# Influence of Perfluorinated Ionomer in PEDOT:PSS on the Rectification and Degradation of Organic Photovoltaic Cells†

Calvyn T. Howells,*[ab] Sueda Saylan,[a] Haeri Kim,[c] Khalid Marbou,[b] Tetsua Aoyama,[d] Aiko Nakao,[d] Masanobu Uchiyama,[de] Ifor D.W. Samuel,[b] Dong-Wook Kim,[f] Marcus S. Dahlem[a] and Pascal André*[bd]

a. Masdar Institute, Khalifa University of Science and Technology, Abu Dhabi, United Arab Emirates. E-mail: calvyn.howells@delaterre.co.uk
b. School of Physics and Astronomy, University of St Andrews, St Andrews, UK.
c. Center for Correlated Electron Systems, Institute for Basic Science (IBS), Department of Physics and Astronomy, Seoul National University, Seoul, Republic of Korea
d. RIKEN, Wako, Saitama, Japan. E-mail: pjpandre@alum.riken.jp
e. Graduate School of Pharmaceutical Sciences, University of Tokyo, Tokyo, Japan
f. Department of Physics, Ewha Womans University, Seoul, Republic of Korea

Poly(3,4-ethylenedioxythiophene):poly(styrenesulfonate) (PEDOT:PSS) is widely used to build optoelectronic devices. However, as a hygroscopic water-based acidic material, it brings major concerns for stability and degradation, resulting in an intense effort to replace it in organic photovoltaic (OPV) devices. In this work, we focus on the perfluorinated ionomer (PFI) polymeric additive to PEDOT:PSS. We demonstrate that it can reduce the relative amplitude of OPV device burn-in, and find two distinct regimes of influence. At low concentrations there is a subtle effect on wetting and work function, for instance, with a detrimental impact on the device characteristics, and above a threshold it changes the electronic and device properties. The abrupt threshold in the conducting polymer occurs for PFI concentrations greater than or equal to the PSS concentration and was revealed by monitoring variations in transmission, topography, work-function, wettability and OPV device characteristics. Below this PFI concentration threshold, the power conversion efficiency (PCE) of OPVs based on poly(3-hexylthiophene-2,5-diyl):[6,6]-phenyl-$C_{61}$-butyric acid methyl ester (P3HT:PCBM) are impaired largely by low fill-factors due to poor charge extraction. Above the PFI concentration threshold, we recover the PCE before it is improved beyond the pristine PEDOT:PSS layer based OPV devices. Supplementary to the performance enhancement, PFI improves OPV device stability and lifetime. Our degradation study leads to the conclusion that PFI prevents water from diffusing to and from the hygrosopic PEDOT:PSS layer, which slows down the deterioration of the PEDOT:PSS layer and the aluminum electrode. These findings reveal mechanisms and opportunities that should be taken into consideration when developing components to inhibit OPV degradation.

## Introduction

Organic photovoltaics (OPVs) are an emerging technology and potential low-cost solution to solar energy conversion. The light-weight and flexible carbon-based semiconductors used in their fabrication have high absorption coefficients, can be solution-processable, and do not require high-temperature processing. Consequently, OPVs and other organic material based electronic devices can be manufactured in bulk using limited material and well-established manufacturing techniques, i.e. ink-jet printing, spin-coating, spray deposition, electrospinning and roll-to-roll processing.[1-14] The interest in developing this organic technology has focussed on the aforementioned aspects, cost and processability, as so far its efficiency and stability are much lower than commercially available inorganic photovoltaics. That said, efficiencies exceeding the 10 % efficiency benchmark required for commercialisation have now been reported.[15-20] This brings to the forefront of research the need to understand the mechanisms responsible for degradation in order to improve the lifetime of the OPV devices. The field is cognisant of those un-encapsulated OPVs lasting days, whereas crystalline Si based solar cells can last several decades.[21]

Tremendous gains in efficiency have been realised using systematic optimisation, facilitated by a deep understanding of material and device science. A similar approach is required to improve the device lifetime. The degradation mechanisms depend on the active layer, interlayer, extraction or charge transport layers and electrode materials, the type of device structures, i.e. regular or inverted, and the environmental conditions during testing. The latter has caused some confusion within the field with different lifetimes being reported for a given c ell. The main reason for this is that organic materials and the electrodes also degrade when exposed to $O_2$ and $H_2O$.[22-29] As a result, lifetime can vary with climate, and even aspects like electrode thickness and deposition rate.[25,30] This has led many researchers to encapsulate their devices or to measure them in an inert atmosphere, even though these different techniques can lead to apparent inconsistencies between laboratories.[31] The lifetime is also sensitive to temperature and illumination conditions.[30-33] Reese *et al.* have published guidelines for testing lifetime to improve the reliability of reported values.[34] Although a number of studies have been carried out and progress made, the aforementioned issues overlap with the inherent complexity of the degradation mechanisms. Many of these are caused or sped up by the ingress of $O_2$ and $H_2O$, and occur concurrently, thus degradation is not yet fully understood. Typical degradation mechanisms in the active layer include photochemical and thermochemical reactions,[35-38] whilst oxidation, delamination and interfacial effects may occur at the electrodes.[39-45] Metal ion diffusion from the electrodes and changes in morphology have also been reported in both the active and poly(3,4-ethylenedioxythiophene):poly(styrenesulfonate) (PEDOT:PSS) layers.[46-53]

PEDOT:PSS is a low-cost, commercially available and environmentally benign polymer blend that is widely used in organic optoelectronics between indium tin oxide (ITO) anodes and active layers.[54-57] With its metal properties,[58-62] PEDOT:PSS and similar conducting layers are part of the contact and are sometimes called hole injection layers in light emitting devices. By analogy, in this solar cell work, PEDOT:PSS is then referred to as a hole extraction layer (HEL) rather than a hole transport layer or interlayer. The large ionisation potential promotes an Ohmic contact and improves the built-in electric field, whilst high conductivity and transparency ensure minimal resistive and optical losses, respectively.[53,63,64] The HEL also helps to prevent metal ion diffusion from the ITO into the active layer and the cathode from short-circuiting the anode.[50] The PSS is a water-soluble polyelectrolyte that serves as a charge balancing dopant during the polymerisation of EDOT monomers. It oxidises and stabilises in aqueous media the otherwise insoluble conjugated polymer PEDOT, leading to a solution-processable blend.[65]

The properties of PEDOT:PSS make it a versatile material and one of the most widely used conducting polymer systems.[53,63-87] Its electronic properties are extremely sensitive to subtle changes in morphology, which has led to a wide range of organic compounds, including acids, alcohols, salts, sugars and surfactants, to be used as co-solvents or used in film post-treatments, resulting in conductivities up to 3000 S.cm$^{-1}$ and work-functions ranging from 4.0 to 5.7 eV.[67-76] This level of conductivity permits the replacement of ITO, a brittle inorganic transparent electrode that uses the scarce element indium, with a flexible, renewable and commercially viable alternative.[84,85] Some of these approaches have produced transparent electrodes with suitable optical and electronic properties for applications in optoelectronics, and ITO-free OPVs with comparable efficiencies have been realised.[77-81] A different strategy was implemented by Stapleton *et al.* increasing the conductivity up to three times that of ITO by using carbon nanotubes and silver nanowires as additives in the PEDOT:PSS thin films.[82,83] K. Leo and co-workers used similar electrodes to fabricate efficient and flexible OPVs.[86]

The film morphology is also sensitive to processing conditions. For instance, annealing has been shown to increase the conductivity and work-function of PEDOT:PSS films by removing residual water content and enlarging the PEDOT-rich domains.[53,67,87] Ultraviolet (UV) irradiated films also show an increase in domain size and a reduction in pores.[63] In organic semiconductors such changes influence charge transport and device performance.[88-90] The increase in domain size is due to conformational changes in the PEDOT. In an untreated film, the dominant structure is benzoid, which favours a coiled conformation, whereas in a treated film the dominant structure is quinoid. The latter favours a linear or expanded coil conformation that increases the conjugation length of the polymer chain.[63]

How co-solvents and post-treatments enhance conductivity has been the subject of much debate. Kim *et al.* proposed that polar solvents with a high dielectric constant such as dimethyl sulfoxide (DMSO), N, N-dimethyl formamide (DMF) or tetrahydrofuran (THF) reduce the Coulombic interaction between the positively charged PEDOT and the negatively charged PSS through strong screening.[91] Whilst O. Inganäs and co-workers suggested that high-boiling-point solvents like glycerol or sorbitol act as plasticisers, improving inter-chain interactions by reorienting the conducting polymer chains.[92,93] Ouyang *et al.* advocated that with ethylene glycol (EG), meso-erythritol and 2-nitroethanol, among others, the increase in conductivity is not driven by the same mechanisms as observed when using co-solvents possessing a high dielectric constant or boiling-point. Rather, those with two or more polar groups influence the morphology much the same as the processing conditions.[68,94]

PEDOT:PSS films are known to have a PSS-rich layer at the surface, and there is evidence to suggest that this insulating layer can be removed with post-treatments,[95-97] if not simply when spin-coating the active layer.[76] Post-treatments usually involve submerging the PEDOT:PSS film in a bath of solvent or spin-coating a solvent on top, examples include but are not limited to: EG, DMSO and glycerol monostearate (GMS). However, such treatments may also, or solely, increase the size of the PEDOT-rich domains.[68,94,96,98]

Unfortunately, PEDOT:PSS is not without its problems. The hygroscopicity, acidity and chemical reactivity can have a detrimental effect on OPV stability and lifetime. This has prompted research into its replacement with transparent conductive oxides (TCOs), such as $MoO_3$, $WO_3$ and $V_2O_5$.[24,99-102] Whilst progress is being made, high temperature vacuum processing and post-deposition annealing are usually required. This is not compatible with the solution-processable, large-area, high-throughput and cost-effective targets of OPVs.

The absorption of water by the hygroscopic PEDOT:PSS alters the morphology and, as a result, the electronic properties in a manner unfavourable for charge extraction, as in OPVs for example, the work-function is moved towards the vacuum level when PEDOT:PSS is exposed to moisture.[53,67] In addition, as many degradation mechanisms are underpinned by the ingress of $O_2$ and $H_2O$, the use of PEDOT:PSS also increases the rate of degradation. An example is the oxidation of the aluminium electrode at the photoactive blend interface in a regular structure.[24,26,103,104] The sulphonic acid groups of PSS can also etch the ITO, leaving products of the reaction in the HEL. Combined to the PSS chemical reactivity, this HEL is then predisposed to be sensitive to the environmental conditions.[50] Whilst the focus of this paper is organic solar cells, our results may also be relevant to conducting polymers used in organic light-emitting diodes.

Conducting layers gain a growing importance for OPVs and the success of the technology relies not only on their optical and electronic properties, but also on their relative lack of sensitivity to $O_2$ and $H_2O$. If this could be achieved with a conjugated system, it would relax the need for encapsulation, and thereby facilitate the fabrication of solar cells matching the technology targets.[24] The ease at which the PEDOT:PSS conductivity and work-function can be tuned combined with its commercially availability and solution-processability would make it an ideal material for optoelectronics, as long as the limitations described earlier can be solved.

Recently, we used anionic fluorinated materials as additives to PEDOT:PSS, to investigate hybrid HELs in two separate photoactive systems.[76] Fluorination leads to processing properties orthogonal to water and oil based materials, which is of interest from multilayer structuration to running chemical reactions.[105-109] In push-pull molecular designs, fluor is also used due to its high electro-negativity.[110] In our OPV work, fluorination was shown to alter not only the work-function of the HEL but also the layer refractive indices, which led to an improvement in charge extraction and photo-generation, respectively.[76] Their relative contribution to the increase in performance was found to be photoactive system dependent. However, both P3HT and PTB7 based systems demonstrated improved device-to-device reproducibility when the HEL contained a fluorinated additive. Our previous study focussed on two additives at a single concentration, which was selected to preserve the conductivity of the HEL.[76] A similar approach with the anionic perfluorinated ionomer (PFI) had proved effective in organic light-emitting diodes (OLED), where it improved both performance and lifetime. The performance enhancement was attributed to superior hole-injection, whilst the increase in lifetime to the blocking of metal ion diffusion.[100,111,112] The latter was also suggested in a recent solar cell work where an increase in lifetime was observed for PCDTBT cells. There, the authors advocated that the ability to tune the work-function of the HEL by altering the relative concentration of PFI could help in achieving a universal HEL that could be used for a wide-range of photoactive systems.[113]

In the present work, we investigate the optoelectronic and topographical properties of the PEDOT:PSS with different PFI concentrations. Then, we show that the work-function, among threshold corresponding to [PFI]≈[PSS]. To demonstrate this, we fabricate and characterise OPVs with poly(3-hexylthiophene-2,5-diyl):[6,6]-phenyl-$C_{61}$-butyric acid methyl ester (P3HT:PCBM). A degradation study is performed in accordance with the ISOS-L-1 protocol.[34] As the overall sulphonic acid group concentration in the HEL thin film is increased with the PFI additive, our findings emphasise that other properties, is not tuned continuously by the PFI concentration. The tuning only occurs above the additive PFI prevents $H_2O$ from diffusing to and from the HEL, rather than being linked to the diffusion of indium and tin ions across the device layers. It is the creation of this moisture barrier with the fluorinated material that is responsible for the improvement in stability and lifetime of the OPV devices.

## Results

### PFI-modified PEDOT:PSS

Information on the materials and the preparation of solutions and films can be found in the Experimental Section of the electronic supplementary information (ESI).† PEDOT:PSS (Fig 1. a1:a2) with varying weight ratios of PFI (Fig 1. b) are labelled as PEDOT:PSS:PFI. Fig. 1c shows the 400 to 800 nm transmittance spectra of composite thin films for different PFI concentrations (Section ESI-1.3.3†).

At low concentrations, i.e. where the concentration of [PFI]<[PSS], e.g. 1:6:1 and 1:6:2.5, the change in transmittance is negligible. Whereas at high concentrations, i.e. where the concentration of [PFI]>[PSS], e.g. 1:6:13.4 and 1:6:30, the decrease in transmittance is larger than the experimental uncertainty. The change takes place nearby the concentration where [PFI]≈[PSS], e.g. 1:6:6, which is confirmed below with other characterisation techniques. The largest difference in transmittance occurs between the thin film with no additive and that with the highest concentration, 1:6:0 and 1:6:30, respectively. This difference also increases with wavelength. The maximum difference is about 2 % at 800 nm. At 500 nm, the difference in transmittance between the lowest and highest PFI concentration thin film is reduced to approximately 1 %. This is illustrated in the inlay of Fig. 1c with the variations at 500 and 800 nm shown for different PFI loadings. The transmittance spectra shows that, regardless of concentration, the PFI content does not alter significantly the optical properties of the thin films. Dektak profilometer measurements (Fig. S2)† show thicknesses of about 32 to 40 nm, which within the experimental uncertainty do not vary much with PFI. Therefore, the PEDOT:PSS:PFI films could be used in the fabrication of OPVs as HELs and compared with one another.

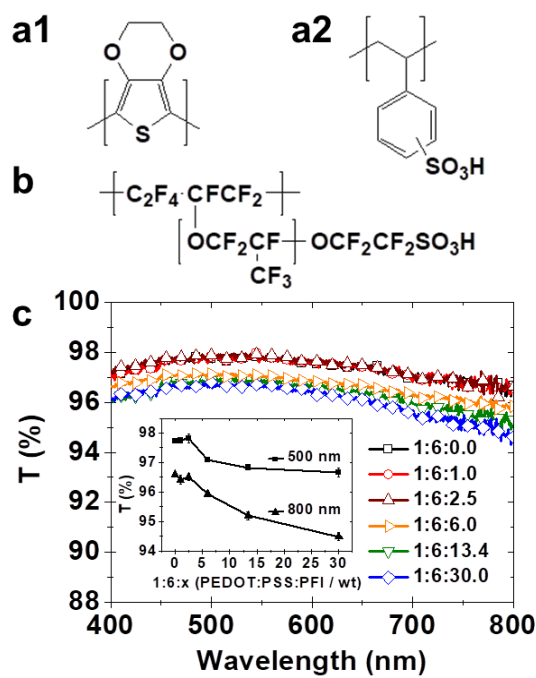

**Fig. 1.** Molecular structures of PEDOT (**a1**), PSS (**a2**), PFI (**b**), and transmission spectra with inlay showing the transmission at 500 nm and 800 nm for different additive weight ratio of PEDOT:PSS:PFI (**c**).

In aqueous solutions, PFI and PSS compete to stabilise and dope the PEDOT polymeric chains. Takahiko Sasaki and co-workers used scattering techniques to reveal that PEDOT:PSS forms micelles with a PEDOT crystalline core that grows when subject to solvent treatments, both pre and post thin film deposition.[69] Such morphology changes can alter the topography of the HEL, which in turn can influence OPV performances.[88,114-118] To assess such an effect with PFI mixing, atomic force microscopy (AFM) was performed (Section ESI-1.3.1†). Fig. 2a shows a set of typical AFM images for the thin films, which are consistent with the literature.[71,77-79] The images look somewhat but not significantly different depending upon the PFI ratio. The contrast between the bright and dark regions originates from a domain (or grain) structure, formed by aggregation, with the lateral size of the bright regions corresponding to the domain size. The dark regions indicate the area between nearby mound-like structures; small-sized grains are also present on these mound-like structures. As the concentration of PFI increases, so does the contrast within the domain structure and the number of small-sized grains. The AFM images acquired at low PFI concentrations, 1:6:1 and 1:6:2.5, differ from that of pristine PEDOT:PSS, 1:6:0, with more peaks approaching 10 nm height. At 1:6:6, there is a change in the contrast and small bead-like structures a few tens of a nanometer in size emerge. There are more of these structures in 1:6:13.4. Some of them amalgamate to form ≥ 100 nm aggregates at 1:6:30, and at this concentration we observe the greatest contrast in the structured domain.

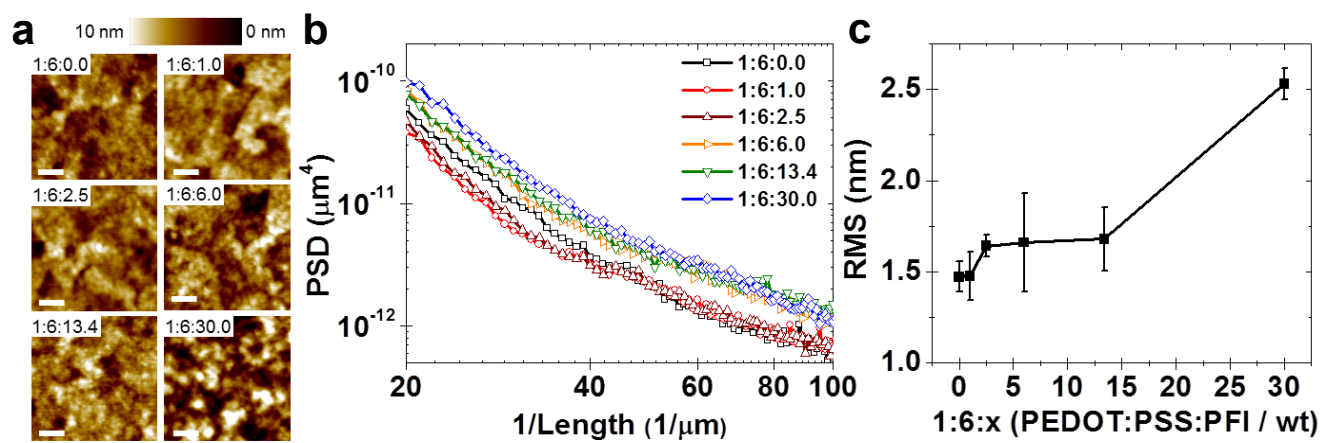

**Fig. 2.** Topography acquired by atomic force microscopy, the scale bar corresponds to 200 nm (**a**), PSD profiles calculated from the AFM measurements for different concentrations of additive plotted (**b**) and RMS as a function of mixing content expressed as weight ratio of PEDOT:PSS:PFI (**c**).

The Fourier transformed and radially integrated topographic images in Fig. 2b suggest further that the presence of PFI does not strongly affect the HEL topography. The power spectral density (PSD) provides the information about the lateral-direction spatial distribution of the height. Thus, the PSD data (Fig. 2b) along with the surface roughness (Fig. 2c) enable us to obtain better understanding of the sample morphology. The PSD analysis is, to some extent, limited by the pixel size and the 1x1 $\mu m^2$ size of the image.[119] In Fig. 2b, it is nonetheless noticeable that PSD at the spatial frequency > 40 $\mu m^{-1}$ is larger for the samples with high PFI concentration, compared with those with low concentration. The three low concentration curves (black-red-brown) pack together below the three high concentration curves (yellow-green-blue) which also pack together in the large spatial frequency region. This suggests that the surface roughness increase for the samples with the high PFI concentration is mainly due to the grains with a few tens of nm size. This is consistent with the RMS values obtained from the topographic images (Fig. 2a) and shown in Fig. 2c and Table S1. The pristine PEDOT:PSS, 1:6:0, and the 1:6:1 concentration share an RMS value of ~1.47 nm. The RMS increases by ~12 % to 1.64 nm for 1:6:2.5. Thereafter, a gradual increase is observed prior to 1:6:30 of 1.66 and 1.68 nm for 1:6:6 and 1:6:13.4, respectively. The highest concentration, 1:6:30, possesses a relatively large RMS value of 2.5 nm in comparison with the other HELs, which is an increase of ~49 % from 1:6:13.4. The uncertainty of topographic measurements and their analysis does not allow to identify reliably a [PFI]≈[PSS] threshold in the HEL in contrast of the transmission spectra data. Overall, PFI alters slightly the topography and leads to higher RMS values but not significantly enough to have any appreciable impact on the OPV performance.

Surface potential maps were simultaneously recorded alongside the topographical images using Kelvin probe force microscopy (KPFM). Fig. 3a (Section ESI-1.3.1†) shows the surface potential relative variation near the average value for each HEL. The relative variations are small, i.e. ±20 mV, and independent of the topography for each sample. The work-functions ($W_{f-KPFM}$) were deduced from the average local surface potentials (Fig. 3a). The results are shown in Fig. 3b and Table S2† alongside those independently acquired with a macroscopic Kelvin probe ($W_{f-mKP}$) in blue and red, respectively. More information on the $W_{f-KPFM}$ and $W_{f-mKP}$ measurements can be found in Section ESI-1.3.† The $W_{f-KPFM}$ and $W_{f-mKP}$ for the pristine PEDOT:PSS film were 4.71 and 5.15 eV, respectively. No significant change in $W_{f-KPFM}$ or $W_{f-mKP}$ was observed for the low concentrations of PFI. The $W_{f-KPFM}$ values were 4.87 and 4.73 eV, and the $W_{f-mKP}$ 5.18 and 5.15 eV, for 1:6:1 and 1:6:2.5, respectively. At 1:6:6, there is a noticeable increase in $W_{f-KPFM}$ and $W_{f-mKP}$. A value of 5.25 eV was found for $W_{f-KPFM}$ and 5.29 eV for the $W_{f-mKP}$. The $W_{f-KPFM}$ for the high concentrations, 1:6:13.4 and 1:6:30, are 5.23 and 5.41 eV, while the $W_{f-mKP}$ are 5.53 and 5.71 eV, respectively. The $W_{f-KPFM}$ and $W_{f-mKP}$ values are slightly different, but follow a similar trend. The values recorded with the mKP are slightly higher than those acquired with the KPFM, which may result from the two sets of data comparing near and far field measurements. The exception is at 1:6:6, where a lower value was recorded by mKP than by KPFM. There is no apparent correlation between the $W_{f-KPFM}$ measurements and the RMS alues, even if the topography images do appear to show a change from 1:6:2.5 to 1:6:6.

To gain further insight into the composite films, wetting properties were investigated using a contact angle goniometer (Section ESI-1.3.4†). The contact angle measurements are dis-

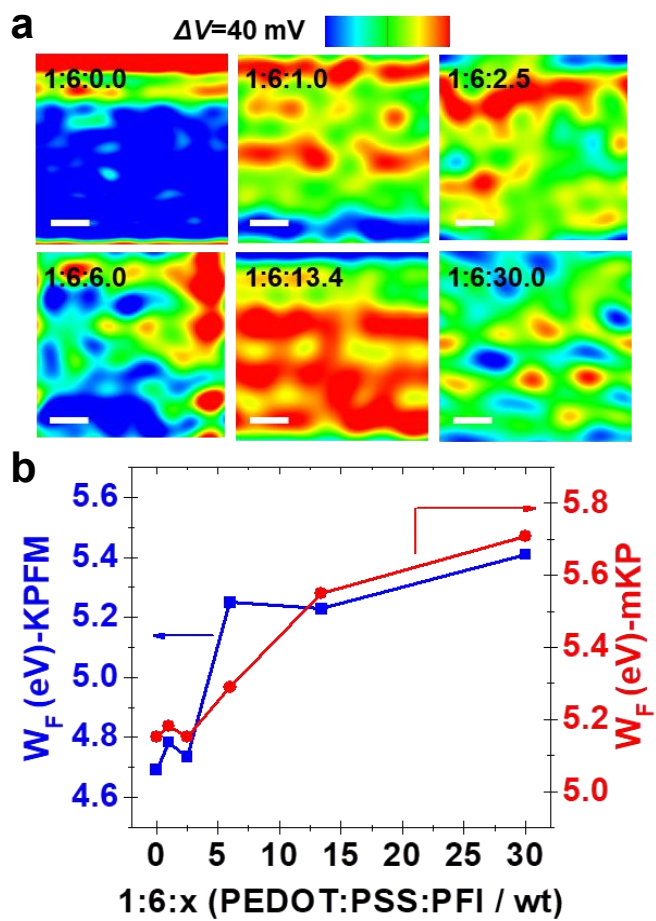

**Fig. 3.** False-colored plot of the relative variation of the surface potential obtained by Kelvin probe force microscopy (KPFM), the scale bar corresponds to 200 nm (**a**). Work-function data measured by KPFM (blue squares, ■) and mKP (red discs, ●) as a function of additive content expressed as weight ratio of PEDOT:PSS:PFI (**b**).

played in Fig. 4a, alongside the calculated adhesion energy using Young and Dupré equations (Section ESI-2.3†), and reveal a similar transition as observed in the transmittance and work-function measurements. At low PFI content, the contact angle and adhesion energy are relatively unaffected by the presence of the fluorinated molecule. For the pristine and 1:6:1 thin film a value of 11.1° and 142.6 mN/m was determined. The uncertainty overlaps with that of the 1:6:2.5 films, where the values are 11.8° and 142.5 mN/m. The threshold that is visible in Fig. 4a corresponds to a concentration where [PFI]≈[PSS]. Here, the contact angle and adhesion energy are 26.7° and 136.3 mN/m, respectively. This is consistent with micelles made of an insoluble PEDOT core surrounded and stabilised by sulfonate molecules. When the amount of PFI is lower than PSS, it mostly competes with it. The relative fraction of PSS replaced by PFI is redistributed between the vicinity of the PEDOT and the surface of the film. Consequently, the contact angle is maintained at a constant value. As the PFI to PSS ratio increases, the added PFI does not find enough room around the PEDOT cores and starts to populate the surface of the film. Naturally, the contact angle is then strongly affected until the PFI molecules saturate the interface leading to a plateau that is visible in the contact angle and the adhesion energy. These are 33.5° and 132.0 mN/m for 1:6:13.4, and 35.9° and 130.3 mN/m for 1:6:30, respectively. A contact angle variation can influence the morphology and thickness of the active layer.[120-122] We have previously shown that the use of PFI in PEDOT:PSS does not alter significantly the thickness or the morphology of the photoactive semiconducting blend layer.[76] The contact angle and adhesion energy results are summarised in Table S3.†

To further confirm these findings, we investigated the change in wettability of the PEDOT:PSS layer due to the PFI additive, via in situ observations of condensation in an environmental scanning electron microscope (ESEM) equipped with a gaseous secondary electron detector (GSED) and a Peltier cooling stage. The sample stage of the ESEM was cooled to 2 °C to enhance the image quality by reducing the vapour density in the specimen chamber, while still obtaining high relative humidity (RH) values. In this case, RH was controlled by varying the water vapour pressure inside the ESEM chamber. The pressure was increased at the onset of 100 % RH, i.e. around dew point (707 Pa and 2 °C),[123] which resulted in condensation on the sample surface. After water droplet formation, the electron beam was switched on, resulting in an increase of the sample's temperature. The water droplets start to evaporate from the sample surface satisfying the equilibrium condition in accordance with the phase diagram of water. Whilst referred to as condensation dynamics, it is the progressive evaporation of the water droplets which is then monitored. SEM frames selected from the movies showing the condensation dynamics are presented in Fig. 4b1-3 and c1-3 for pristine PEDOT:PSS and PFI modified PEDOT:PSS at 1:6:30, respectively. As seen from Fig. 4b1 and c1, the water droplets nucleated on the pristine PEDOT:PSS film with significantly lower contact angles than on the PFI modified PEDOT:PSS. This indicates a decrease in the wettability of the surface with the addition of PFI polymer, which is consistent with the characterisation completed in air and presented in Fig. 4a. Fig. 4b2-3 and c2-3 show the gradual evaporation of the condensed droplets from the two samples, respectively, as a result of the heating of the sample surface by the electron beam.[124] The higher wettability of the pristine PEDOT:PSS surface is also apparent in the evaporation dynamics; the distinct droplets observed in Fig. 4b1 elongate into the film like structures on the surface that remain visible after 50 seconds (Fig. 4b3). On the contrary, the projected area of the individual droplets initially formed on the surface of the PFI modified PEDOT:PSS are maintained from condensation to full evaporation of the droplets, which lasts only 15 seconds, as seen from Fig. 4c1 to c3. It is also apparent that there are PFI and PSS rich regions on the surface of the 1:6:30 HEL, as some areas absorb water while others do not, as shown in Fig 4c1.

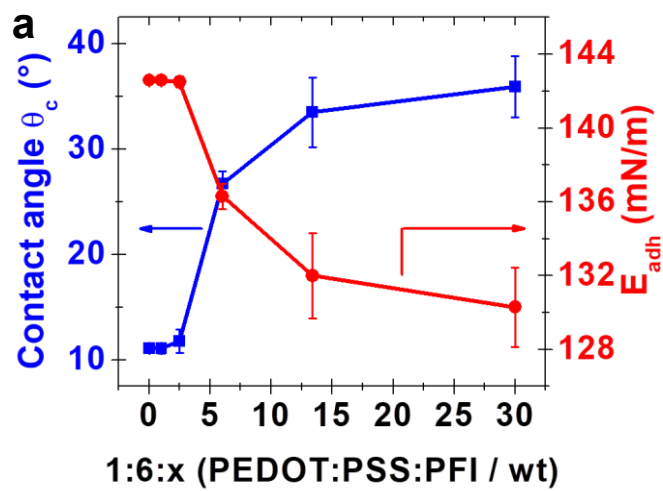

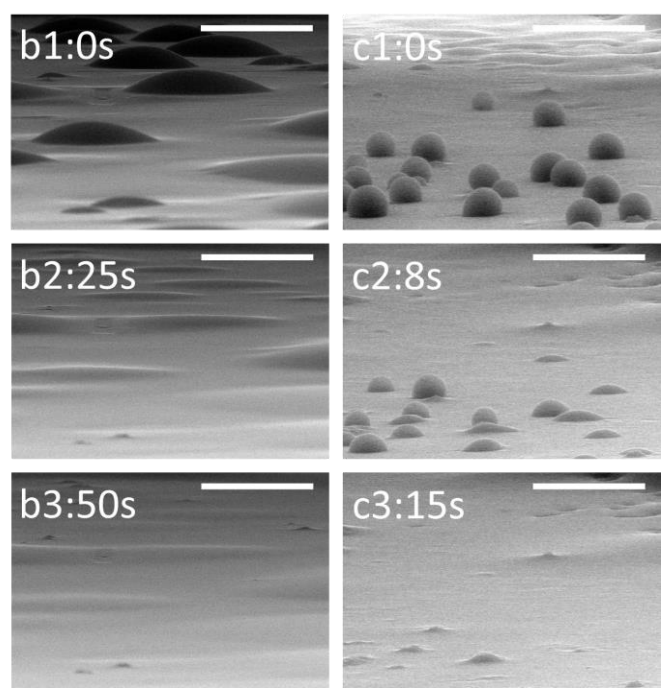

**Fig. 4.** Contact angle (blue squares, ■) and adhesion energy (red discs, ●) as a function of the additive content expressed as weight ratio of PEDOT:PSS:PFI (**a**). Environmental scanning electron microscopy images (×3500, scale bar 30 μm) of pristine PEDOT:PSS (**b**) and PEDOT:PSS with PFI (**c**). Frames selected from the movies in ESI showing early (**1**), intermediate (**2**) and final stages of contact angle (**3**). The accelerating voltage of the electron beam, the vacuum, the temperature were set to 20 kV, 860 Pa and 2 °C, respectively.

To summarise, investigating the PEDOT:PSS:PFI composite material revealed that, while the topographical properties are not significantly altered from an OPV device point of view with PFI the optical, electronic and wetting interfacial properties can be tuned when PFI concentration reaches a threshold. That is, however, not always the case, as tuneability is only observed above a specific PFI concentration, which we assume depends on the initial ratio of PEDOT and PSS in the commercial preparation. Because of the identified threshold, that is [PFI]≈[PSS], our results are dissimilar to, but complement those in the literature.[100,112,113]

### Rectification of OPVs

The remainder of the study focuses on P3HT:PC$_{61}$BM bulk heterojunction OPVs fabricated with PEDOT:PSS:PFI composite HELs. Information on the materials can be found in Section ESI-1.1.† The chemical structures of the photoactive materials are presented in Fig. 5a. The OPVs were fabricated using a regular non-inverted structure with ITO as the anode. The HEL followed by the photoactive blend were spin-coated on-top of the ITO anode, and an aluminium cathode was deposited perpendicular to it by thermal evaporation. The device structure is presented in Fig. 5b. Further fabrication and characterisation details can be found in Section ESI-1.2.†

The flat band energy level diagram presented in Fig. 5c illustrates the charge extraction pathways. The ability to transport charge between materials is related to a favourable and close alignment of the donor and acceptor states, which are a molecule's occupied and unoccupied frontier orbitals, respectively. The PEDOT:PSS -4.9 eV literature value is comparable to the Kelvin probe data reported in Fig. 3b for the pristine HEL film.[63,100,102,112,125] It is known that the energy level

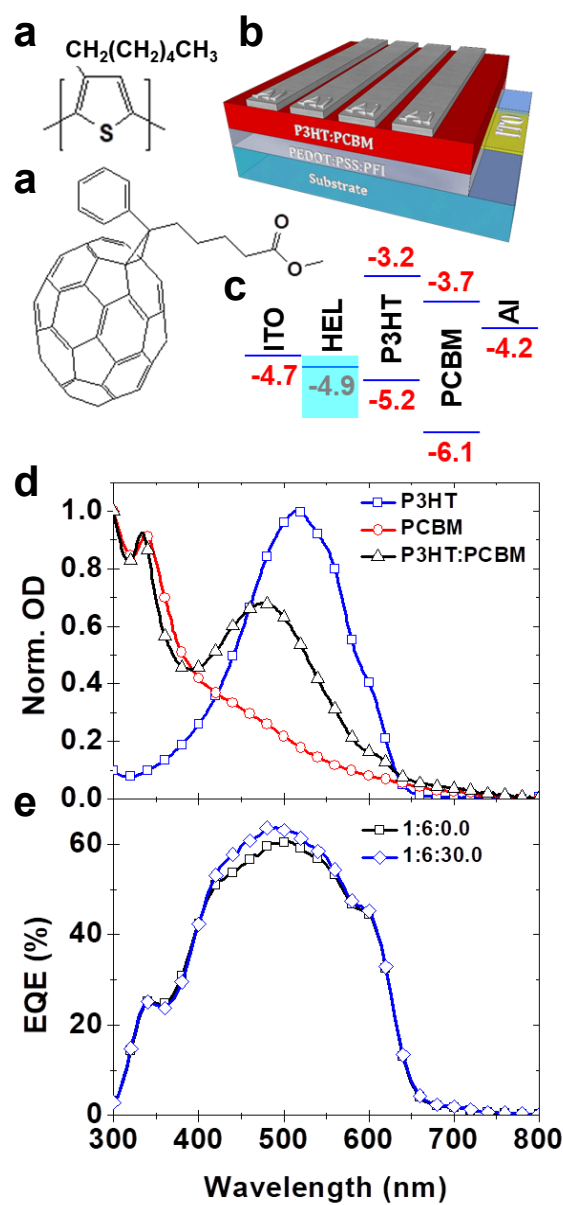

**Fig. 5.** Molecular structure of P3HT (**a1**) and PCBM (**a2**). Schematic of the devices (**b**) and flat band energy level diagram The turquoise box shows the range of workfunctions obtained as the PFI content of the conducting polymer was varied (**c**). Normalised absorbance for P3HT (blue squares), PCBM (red discs) and P3HT:PCBM (1:1) (black triangles) (**d**). External quantum efficiency (*EQE*) for P3HT:PCBM with (blue diamonds) and without additive (black squares) (**e**).

 

is shallow, but closer to the HOMO of P3HT than the ITO. This shallowness increases the *PCE* by improving the charge selectivity, contact resistance and built-in field.[113] PEDOT:PSS not only enhances charge extraction, but also improves device lifetime by preventing elements diffusing from the ITO into the active materials.[50,126,127] However, the PEDOT:PSS layer is hygroscopic, and on exposure to $O_2$ and $H_2O$, the work-function ($W_f$) is reduced towards vacuum.[53,75] A favourable Ohmic contact requires the energy level to be ≤ -5.2 eV.[128,129] Therefore, our interest in PFI is twofold: first to encapsulate the PEDOT layer and so prevent $O_2$ and $H_2O$ from diffusing to and from it; second to increase the work-function of the conducting polymer layer for easier and faster hole extraction thanks to an improved contact and interfacial charge transfer between the HEL and the P3HT located nearby.[76]

The semi-transparent turquoise box in Fig. 5c illustrates the HEL work-function range covered when varying the PFI amount, as measured by KPFM and displayed in Fig. 3b. The increase in work-function of PEDOT:PSS with PFI was first reported in OLEDs, where it was found to improve the efficiency and lifetime with a supposed work-function gradient and a metal ion diffusion barrier, respectively.[100,111,112] To our knowledge, the appearance of PFI in OPVs was to help understand the morphology of PEDOT:PSS by studying P3MEET:PFI via neutron reflectometry.[130] Subsequent studies have led to similar conclusions as with the OLEDs.[113,125]

Fig. 5d presents the normalised absorbance spectra of the semiconductors used in this investigation. Pristine P3HT covers a 400 to 650 nm wide spectral range, while the fullerene derivative, $PC_{61}BM$, presents a high energy peak followed by a relatively long and monotonous tail. The features of both materials are noticeable in the absorbance of the blend formed with a 1:1 weight ratio, as used for our devices.

Fig. 5e shows typical external quantum efficiencies (*EQE*) of devices with two different conducting polymer layers. The shape of the *EQE* spectrum of the pristine PEDOT:PSS based device is consistent with the literature.[131] As expected, the *EQE* of the pristine PEDOT:PSS device covers a spectral range similar to the P3HT:$PC_{61}$BM absorbance spectra. The *EQE* peak that is around 500 nm is a little over 60 %, and displays a kink at high energy which is consistent with the composition of the blend. The shape of the *EQE* spectrum is not drastically altered, but a slight increase in the *EQE* is nonetheless observed with PEDOT:PSS fluorination. At 450 nm, the *EQE* values differ by about 8 %. This is driven by an increase of the optical electric field intensity distribution within the device, finding its origin in the variation of optical constants induced by fluorination led charge transfers.[76] High efficiency OPVs usually consist of low band-gap polymers, but these tend to have issues with stability and scalability.[132] P3HT is relatively stable, even when exposed to moisture, and readably scalable thanks to its straightforward synthesis.[132-134] It is also compatible with high-throughput production techniques and is the most studied and understood polymer in the field. State-of-the-art P3HT OPVs have been shown to achieve efficiencies in excess of 7 %.[135] This guided our choice of active materials. Fig. 6 shows the performance parameters for the studied PEDOT:PSS:PFI ratios. Short-circuit current density ($J_{SC}$), open-circuit voltage ($V_{OC}$), fill factor (*FF*) and power conversion efficiency (*PCE*) presented in Fig. 6 were extracted from the *J-V* characteristics in Fig. 7a and Fig. S4. Un-fluorinated PEDOT:PSS based devices display efficiencies and characteristics consistent with those reported in the literature.[76,113,130] However, we can see that varying the PEDOT:PSS:PFI ratio has a non-monotonous effect on the device characteristics, which is somewhat different from the literature,[100,111-113,136] but in agreement with our composite film studies. Previous reports have shown a gradual increase in work-function and device performance with PFI content, and advocate the existence of a concentration gradient with the upper surface of the PEDOT:PSS being rich in PFI.[100,111-113,130] At low concentrations of PFI, there is a significant decrease in the $J_{SC}$ as shown in Fig. 6a. The $J_{SC}$ recovers with increasing further the amount of PFI to eventually where the data displays a relatively low variation between the highest PFI concentration and the pristine HEL. This suggests that the number of photo-generated charges is not drastically altered by the HEL fluorination. This is confirmed by calculating the $J_{SC}$ from the *EQE* in Fig. 5e and taking the difference, which is ~0.2 mA/cm$^2$, and is also in agreement with the literature for P3HT:PCBM bulk heterojunction OPVs.[76,113]

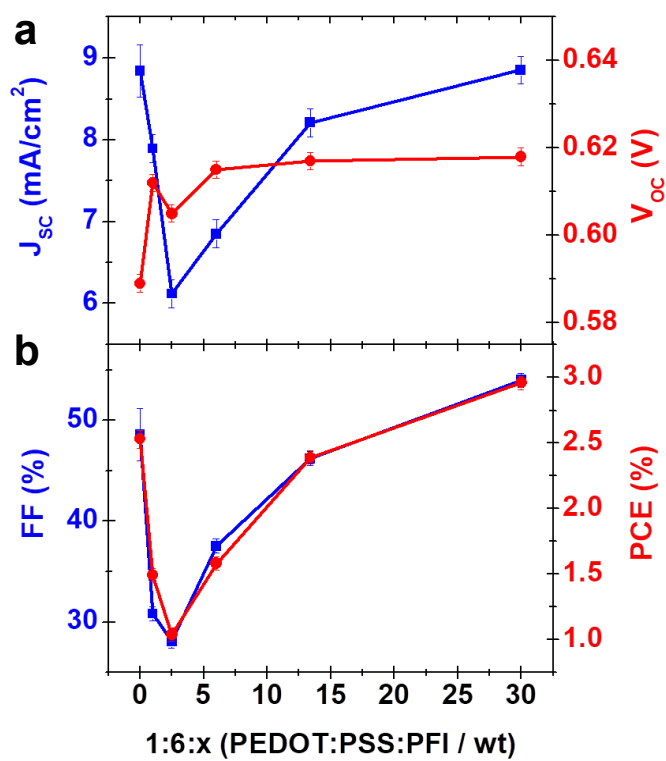

**Fig. 6.** Variation of the short-circuit current density ($J_{SC}$) and open-circuit voltage ($V_{OC}$) (**a**), fill factor (*FF*) and power conversion efficiency (*PCE*) (**b**) of P3HT:PCBM photovoltaic cells as a function of PFI content in composite HEL films expressed as weight ratio of PEDOT:PSS:PFI.

Fig. 6a also shows that the $V_{OC}$ increases progressively with fluorination until it reaches a plateau. The only exception to this trend is at 1:6:2.5, where the $V_{OC}$ is smaller than that of 1:6:1 and 1:6:6, but remains larger than for 1:6:0 based devices. The fluorination of PEDOT:PSS translates into a ~30 mV maximum variation, which is the result of a favourable shift of work-function that PFI induced in the HEL and associated with a decrease in charge carrier recombination within the device. This is consistent with an alteration of the band bending and energy alignment within the devices. As described by the Integer Charge Transfer model,[137] it creates surface dipoles and an internal electric field at the HEL/bulk heterojunction interface, which not only sweeps out charges and prevents electrons from recombining at the hole-extracting electrode, but also reduces charge carrier relaxation and recombination at the donor acceptor interface within the bulk heterojunction.[76,138-140] It is noteworthy that at low PFI concentrations the $V_{OC}$ is increased compared with PEDOT:PSS, when there is no significant change in the work-function. It is possible that the presence of low PFI concentration in the HEL and the resulting S-kink hinder access to the intrinsic $V_{OC}$ of the devices. Nonetheless, the variation in the apparent $V_{OC}$ is small but, in this concentration range, the deformation in the *J-V* characteristics decreases the reliability of the $V_{OC}$ which can be extracted from these curves. The *FF* is an important performance parameter that reflects the efficiency at which charges are extracted out of the device. As seen in Fig. 6b, the *FF* exhibits a similar response as the $J_{SC}$ and consequently the *PCE*. The *FF*, $J_{SC}$ and *PCE* for the pristine HEL are 48.6 %, 8.84 mA/cm$^2$ and 2.53 %, respectively. These values decrease at low PFI content. The lowest efficiency is 1.04 % at 1:6:2.5 with a *FF* and $J_{SC}$ of 28.1 % and 6.12 mA/cm$^2$, respectively. At 1:6:6 where PFI≈PSS, we begin to recover the performance. Here, the *FF*, $J_{SC}$ and *PCE* are 37.6 %, 6.85 mA/cm$^2$ and 1.58 %, respectively. The device with the highest efficiency occurs at 1:6:30 and possesses a *FF*, $J_{SC}$ and *PCE* of 54.0 %, 8.85 mA/cm$^2$ and 2.96 %, respectively. It is worth mentioning that as the PFI increases beyond the threshold, the recovery of the device characteristics (Fig. 6b) occurs when the added PFI slightly reduces the transmission of the HEL (Fig. 1c). Given the loss in transmission, it is then remarkable that the high PFI devices are better than the pristine ones.

It is also noticeable that at low but not null additive amount, the decrease in device performance is associated with an S-kink in the *J-V* characteristic which reduces the *FF*. The origins of S-kinks are usually considered to be due to at least one of the following: i) interfacial dipoles or defects, ii) charge carrier mobility imbalance and iii) energy barriers.[87,141-145] The equivalent circuit model based on elementary circuit elements is used to understand the loss mechanisms that influence the *FF* at low PFI concentrations. In the present study, these losses are largely related to the series ($R_S$) and shunt ($R_{SH}$) resistances, as the composition and morphology of the active layer, and the temperature are unchanged. Contact resistances between the active layer, interlayers, charge transport layers, and electrodes, and the bulk resistances of the above are usually responsible for $R_S$; whilst the parameters that influence the current leakage and the recombination of charge carriers, such as the active layer thickness, illumination intensity and interfacial morphology are impacting on $R_{SH}$.[146] Both $R_S$ and $R_{SH}$ limit the *FF*, as shown in Equation S16.† When no losses are present, $R_S$ and $R_{SH}$ are equal to 0 and ∞, respectively, and one recovers an equation for an illuminated photovoltaic cell, where the response is diode dependent (Equation S15†). At a PEDOT:PSS:PFI ratio of 1:6:2.5, PFI induces a low $R_{SH}$ and a large $R_S$ of 0.334 kΩ.cm$^2$ and 52.3 Ω.cm$^2$, respectively (Table S4†). An $R_S$ of 5.41 Ω.cm$^2$ and a $R_{SH}$ of 0.671 kΩ.cm$^2$ were recorded for the pristine HEL.

The low $R_{SH}$ at 1:6:2.5 is responsible for a significant current leakage, which is also present in the dark-current measurements shown in Fig. S4.† At 1:6:6, i.e. [PFI]≈[PSS], the dark-current is reduced. The $R_{SH}$ values for low concentrations of PFI are less than those obtained at 1:6:0, and for concentrations greater than or equal to 1:6:6, $R_{SH}$ values are larger than observed at 1:6:0. The $R_S$ increases with additive until 1:6:2.5, after which it decreases. The $R_S$ values tend to be larger when PFI is present. The equivalent circuit model shows that $R_S$ reduces $J_{SC}$ and not $V_{OC}$, whereas $R_{SH}$ reduces $V_{OC}$ and not $J_{SC}$. Therefore, it is the $R_S$ and not $R_{SH}$ that is responsible for the deterioration of the device performance. The low $R_{SH}$ at 1:6:2.5 does influence the performance, by leading to a slight decrease in $V_{OC}$. However, it is the S-kink that is responsible for the low $R_{SH}$ at low concentrations of PFI. This is indirectly due to the large $R_S$ value, i.e. poor charge extraction, which increases charge carrier accumulation, i.e. recombination. The most probable reason for this behaviour is that at low concentrations, the PFI disrupts the PSS shell as well as charge transport between the PEDOT grains, leading to a detrimental impact on device performance. As the concentrations increases to reach [PFI]≈[PSS], aggregates of PFI form at the surface of the HEL, where its electronegativity compensates for the above mentioned charge transport disruption, improves charge extraction and consequently device performance. Disrupting charge transport between the PEDOT grains is consistent with an increase in the resistivity of the PEDOT:PSS based HEL, i.e. $R_S$. A detailed description of the equivalent circuit analysis can be found in ESI-3.1.†

**Degradation of OPVs**

We next investigated the effect of PFI on the stability and burn-in of organic solar cells. To provide a more reliable product, manufacturers tend to complete a burn in step on electronic devices presenting a fast initial failure. In this context, the OPV initial and fast degradation is referred to as "burn-in" time. It is often characterised by a drastic and almost exponential PCE decay, which is followed by a slower and more linear deterioration of the device power conversion efficiency. Identifying the physical mechanisms involved in the degradation during the device early stages of operation is essential but made difficult by the fact that the burn-in magnitude and duration vary from one OPV system to another.[147-151] Fig. 7 presents *J-V* characteristics of four selected PFI concentrations and three characterisation times. The performance parameters are shown in Table 1. The transition region, revealed in all our measurements between $x$ = 2.5 and 6, was avoided to obtain a reliable analysis. The degradation study was thereby limited to OPVs with similar $R_S$ and $R_{SH}$ values, i.e. at low and high PFI content, e.g. 1:6:x with $x$ = 1, 13.4 and 30. Between 1.0 and 0.5 V, all the *J-V* characteristics display a bell shape, from which the $V_{OC}$ value is deduced. The $V_{OC}$ increases with PFI concentration. The value for the pristine HEL is 589 mV, and 612, 617 and 618 mV for $x$ =

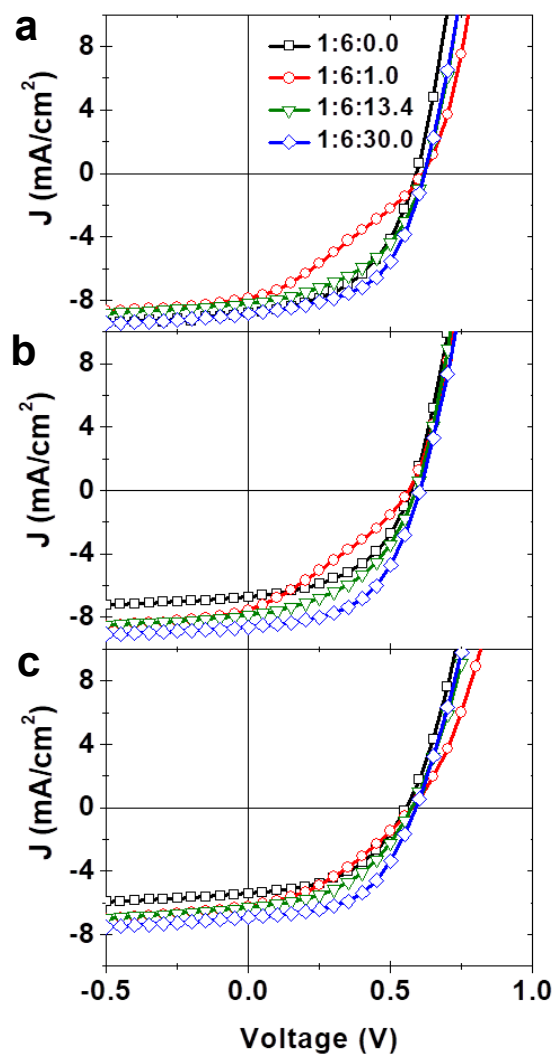

**Fig. 7.** Device characteristics for P3HT:PCBM photovoltaic cells with different concentrations of PFI additive in the PEDOT:PSS based HEL at 0 h (**a**), 0.5 h (**b**) and after 24 h of continuous AM1.5G illumination at 100 mW/cm$^2$ (**c**).

**Table 1** Statistics for P3HT:PCBM device characteristic s at 0, 0.5 and 24 hours. PFI loading of PEDOT:PSS expressed in weight ratio 1:6:x.

| PFI | $J_{SC}$ (mA/cm$^2$) | $V_{OC}$ (mV) | FF (%) | PCE (%) |
|---|---|---|---|---|
| 0.0 | 8.84/6.72/5.43 | 589/570/555 | 48.6/48.1/47.1 | 2.53/1.85/1.42 |
| 1.0 | 7.89/7.54/6.23 | 612/562/574 | 30.8/31.1/36.8 | 1.49/1.32/1.32 |
| 13.4 | 8.21/7.86/6.30 | 617/588/572 | 47.2/46.2/45.8 | 2.39/2.13/1.65 |
| 30.0 | 8.85/8.63/6.93 | 618/602/587 | 54.0/52.7/52.9 | 2.96/2.74/2.15 |

1, 13.4 and 30, respectively. Decreasing the external bias, a change of slope is noticeable around 0.5 V, with a slower decay, which reaches a plateau near -8 mA/cm$^2$ for a small positive external bias. The $J_{SC}$ values for 1:6:0, 1:6:13.4 and 1:6:30 are 8.84, 8.21 and 8.85 mA/cm$^2$, respectively. A slight S-kink in 1:6:1 results in a lower $J_{SC}$ of 7.89 mA/cm$^2$. The feature is also responsible for a FF of 30.8 %, compared to 48.6, 47.2 and 54 % for PFI concentrations of 1:6:0, 1:6:13.4 and 1:6:30, respectively. Under reverse bias, the current density decreases monotonously until it reaches a maximum at -0.5 V between 8.68 and 9.48 mA/cm$^2$, depending on the PFI amount present in the HEL. The PCE calculated from the above-mentioned for 1:6:30 was 2.96 %, ~ 15 % larger than the 2.53 % value acquired with the pristine HEL. The lowest PCE in our degradation analysis is 1.49 % at 1:6:1. Although, the fill factor and $J_{SC}$ begin to recover at 1:6:6, the values for 1:6:13.4 are still lower than those of both the pristine PEDOT:PSS and 1:6:30 HELs, which result in a PCE of 2.39 %.

The degradation measurements were carried out in accordance with the ISOS-L-1 protocol, measured under ambient environmental conditions and continuous AM1.5G illumination with a solar simulator, as described in Section ESI-1.2.2.[34]† Temperature and relative humidity were also measured with the data being shown in Fig. S10.† The mean temperature and relative humidity were found to be 24.94 ± 0.25 °C and 44.62 ± 0.87 %, respectively. After 30 min of operation, the shapes of the J-V characteristics shown in Fig. 7b are similar to those obtained initially and presented in Fig 7a. However, in Fig 7c it can be seen that asoperation time passes the S-kink of the low PFI concentration devices disappears, suggesting a change of the interfacial resistance and more balanced charge extraction.

Fig. 8 presents the normalised variations of the device performance parameters as a function of time for the same four PFI concentrations as in Fig. 7. As shown in Fig. 8a, the additive free device is characterised by a sharp ~ 25 % collapse in $J_{SC}$ within the first half-hour of operation. After this, the $J_{SC}$ continues to decrease with a softer and almost linear decay of ~ 0.0542 mA.cm$^{-2}$/h. The dramatic change in kinetics suggests that the initial degradation mechanism has been exhausted and that another has taken over. The $J_{SC}$ is about 40 % of its initial value after 24 h. For the 1:6:1 PFI composite, the $J_{SC}$ also presents a sharp decrease during the first half-hour, which however only amounts to ~ 5 % of the initial $J_{SC}$ value.

As for the pristine HEL based device, the $J_{SC}$ then continues decreasing almost-linearly to lose ~ 20 % of the initial value after 24 h of operation. For higher PFI amounts, the $J_{SC}$ initial sharp decay becomes barely noticeable and its amplitude is only of the order of a few percent. As the PFI molecules saturate the HEL interface when [PFI]>[PSS], the variation of $J_{SC}$ over time is mostly characterised by the relatively linear slow decay. This slow decay is of the order of 0.0797 ± 0.0004 mA.cm$^{-2}$/h, i.e. less than 1 % per hour. For these PFI composite HEL devices, the initial degradation mechanism is apparently still on-going after this 24 h operation which is in strong contrast with devices based on pristine PEDOT:PSS HEL.

In Fig. 8b, $V_{OC}$ presents a similar behaviour for the pristine HEL and those prepared with a conducting polymer layer containing more PFI than PSS. Its dynamic is characterised by an almost exponential decay, which, however, remains relatively small. After 24 h of operation, the $V_{OC}$ has indeed only dropped by ~ 6 ± 1 % when compared with its initial value, corresponding to an average loss rate of ~ 1 mV/hour. The pristine HEL based device displays the fastest $V_{OC}$ decay. More interestingly, the 1:6:1.0 $V_{OC}$ variation with time stands out. In the initial 30 min, it abruptly decreases by ~ 8 %and then recovers slightly with a 2 % increase. This is discussed below in relation with the FF variation shown in Fig. 8c. The theoretical maximum value of $V_{OC}$ is the difference between the HOMO of the donor and the LUMO of the acceptor, with voltage loss at the electrodes and several other limiting factors having been identified.[152,153] It is the loss at the electrodes which is of relevance here. It is indeed likely that the formation of an oxide layer on the aluminium cathode, in combination with a poor performing anode, leads to more balanced charge extractions and fewer recombinations within the cell. These explain the increase in $V_{OC}$ and *FF* over time at this 1:6:1.0 PFI concentration.

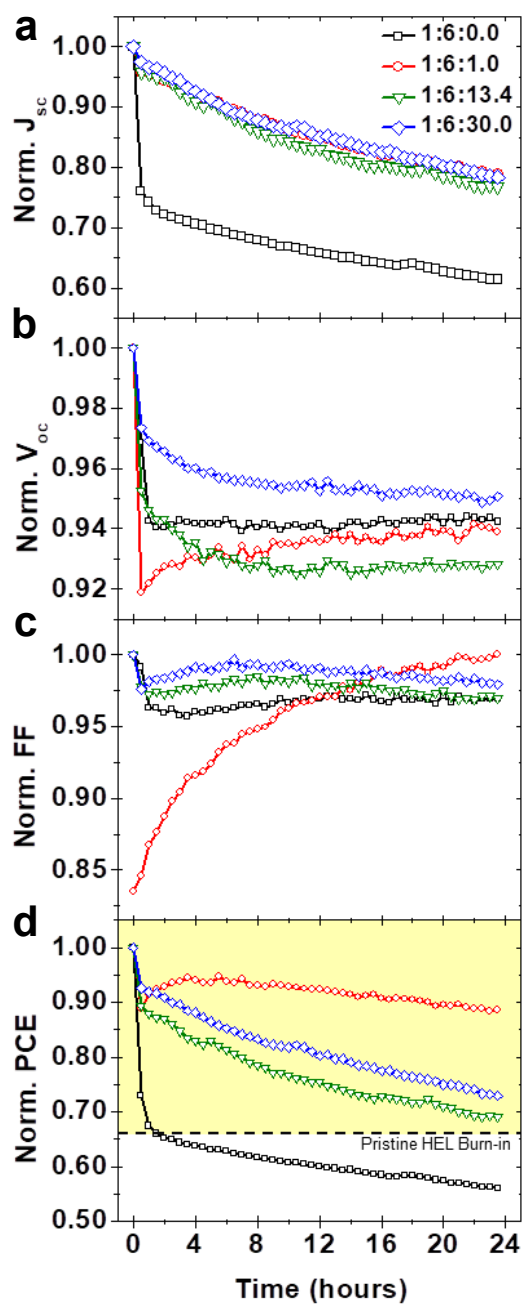

**Fig. 8.** Degradation study with measurements performed under continuous AM1.5G illumination at 100 mW/cm² (data recorded at 30 min time intervals). Normalised short-circuit current density ($J_{SC}$, **a**), open circuit voltage ($V_{OC}$, **b**), fill factor (*FF*, **c**) and power conversion efficiency (*PCE*, **d**) of P3HT:PCBM photovoltaic cells for four PFI additive contents in the composite HEL : without (black) and with additive (red, olive and blue corresponding to weight ratios of *x* = 1.0, 13.4 and 30.0, respectively). The normalisation is based on the first characterisation data except for *FF* with *x* = 1.0 which uses the last due to the presence of the S-kink, see text for details.

More generally, *FF* displays a relatively weak variation at both pristine and high PFI concentrations. The pristine HEL based devices lose ~ 3.5 % of the initial *FF* value after 24 h, which is comparable to what is observed for the HEL made of 13.4 and 30.0 in weight of PFI additive. In contrast, the *FF* of the devices with [PFI]<[PSS] based HELs are characterised by a low initial *FF* value due to the S-kink discussed in Fig. 6b and Fig. 7. However, Fig. 8c shows that, instead of decreasing further with operation time, *FF* of the S-kinked devices increases by more than 25 % over the time scale of the investigation. For readability purposes, the normalisation of the 1:6:1.0 *FF* curve in Fig. 8c is then based on the final rather than the initial *FF* value. Noticeably, both $V_{OC}$ and *FF* recover for the 1:6:1.0 and 1:6:2.5 PFI concentrations devices, which matches with the S-kink disappearance from the *J-V* curves.

Finally, Fig. 8d displays the variation of the normalised *PCE* as a function of time, and the effect of the PFI additive on the burn-in is striking. The *PCE* of the pristine PEDOT:PSS HEL devices decreases sharply by about ~35 % in the first 1.5 h of operation corresponding to the burn-in region. It is illustrated in Fig. 8d with a yellow box and black dashed line, outside of which the *PCE* decays almost linearly with an average loss rate of 0.40 %/hour. After 24 h of operation, the *PCE* of these devices has fallen by ~44 % when compared to the initial characterisation.

In contrast, the devices containing PFI experience a shorter burn-in time of about 30 min which is associated with a *PCE* decrease a lower amplitude and ranging only from 7 to 12 %. For the first 30 min of operation, a 10 % decay occurs with the 1:6:1.0 devices, which is followed by a 5 % recovery. It is unusual for a post burn-in region but it is consistent with the *FF* and $V_{OC}$ behaviour of the low PFI concentration devices. After 4 h operation, *PCE* decreases again, but rather slowly: it only loses 11 % of its initial value after 24 h of operation with an average loss rate of 0.30 %/h. At this low PFI concentration, the S-kink disappearance compensates and partially hides the burn-in mechanism. At high PFI concentration and for the first 30 min of operation, both 1:6:13.4 and 1:6:30.0 devices present an early fast degradation of ~ 11 and 8 %, respectively. It is followed by an almost linear degradation of -0.80 %/h, as expected from a post burn-in regime. Noticeably, the burn-in region presents a much smaller relative amplitude as the PFI additive amount is increased in the HEL. The 1:6:13.4 devices present a 31 % *PCE* decay within 24 hours of operation, in comparison to ~ 27 % for the 1:6:30.0 devices. At these PFI concentrations, the fluorinated additive molecule reduces the relative amplitude of burn-in region by up to ~140 and 167 %, respectively. Incidentally, it is also noticeable that the devices fabricated with the fluorinated additive in the HEL systematically present lower standard variations of the *PCE*, *FF* and $J_{SC}$.[76] In view of the present results, this stabilisation is attributed to the fluorinated nature of the additives and the control it offers on the burn-in degradation mechanism.

## Discussion

The previous literature suggests the existence of a work-function gradient in accordance with a concentration gradient in the PEDOT:PSS:PFI films, and that PFI mainly resides at the surface of the HEL.[100,111-113,136] In the present study, it is revealed that there is a concentration threshold, [PFI]≈[PSS], above which PFI can be found at the surface of the HEL. When [PFI]<[PSS], PFI competes with PSS to stabilise the PEDOT. Around the threshold concentration, the excess of PFI starts to aggregate and populate the surface of the film. Therefore, for one of the most commonly used PEDOT:PSS formulations, Heraeus Clevios™ AI 4083, the work-function is not continuously tuned with the PFI concentration, and the existence of a gradient work-function is unlikely. The above-mentioned is confirmed with the characterisation of OPVs.

The S-kink in the *J-V* characteristics at low concentrations of PFI severely reduces the *FF* and thereby the device efficiency (*PCE*). As previously mentioned, the appearance of this feature usually results from interfacial dipoles or defects, a charge carrier mobility imbalance and/or energy barriers. Wagner *et al.* showed that the inter-diffusion of metals into organic semiconductors can act as recombination centres and result in S-kinks.[87] This type of behaviour may also result from oxidation of an electrode,[49,102] as further discussed below.[50,103]

Other factors, which can influence charge carrier transport, include, but are not limited to, the active layer composition, morphology and thickness.[87,154-156] Tress *et al.* demonstrated also that increasing barrier heights can lead to S-kinks in the *J-V* characteristics.[157] In our case, the low PFI concentration disrupting charge transport between the PEDOT grains could result in charge carrier accumulation, which is responsible for the S-kink in the *J-V* characteristics of these devices. At high concentrations, the excess of PFI compared to PSS can then be found both around the PEDOT grains and at the surface of the HEL. This is consistent with the observed work-function variation (Fig. 3b) which improves the charge extraction efficiency. The charge extraction improvement increases the *FF* (~ 10 %) and $V_{OC}$ (~ 5 %) resulting in the observed *PCE* enhancement (~ 15 %) for the devices with the highest PFI concentration.

Regarding device stability, there have been suggestions that PFI improves OLED and perovskite PV device lifetime by preventing the acidic PSS from etching the ITO and/or blocking the diffusion of metal ions into the organic semiconductor materials.[112,113,158] This is nonetheless questionable, as the PFI possesses the same sulfonic acid group as PSS (Fig. 1a). However, if this were to be true, a significant decrease in *FF* would be apparent, as demonstrated with nanoparticles or isolated metal islands based systems.[120,159-162] However, the decrease in *FF* between the initial and final measurement in the present system is only of ~ 3 % for all the PFI concentrations but 1:6:1, which shows a strong increase of ~ 20 % as presented in Fig. 8c. For this blend, it is possible that a decrease in charge extraction efficiency due to oxidation or delamination of the Al cathode, could result temporarily in a more balanced charge extraction with operation time and therefore incidentally recover part of the *FF* initially lost in the S-kink.[39-43] An alternative explanation is that, while characterising the devices, there is a progressive increase in the HEL conductivity and/or formation of a dipole layer at the interface due to the PFI molecules. This could occur through the water molecules brought in by the hygroscopic sulfonic acid groups altering the HEL morphology and counteracting the disruptive charge transport at low concentrations of PFI between the PEDOT grains.[53,57,163,164] Such a morphology change during the device characterisation would also explain the progressive increase in $V_{OC}$ for 1:6:1 (Fig. 8b).

Metal ion diffusion would also significantly reduce the $V_{OC}$. However, the decrease between the first half-hour and the end of the measurements for all concentrations is between 2-3%. Therefore, if metal ion diffusion were involved, a substantial amount of them would have to enter the organic semiconductors in the first half-hour of operation. Noticeably, the devices containing PFI also experience a loss in $V_{OC}$, consequently it is unlikely that PFI blocks metal ions from diffusing into the organic semiconducting materials. In the present work, it is the $J_{SC}$, but not the *FF* and $V_{OC}$, which is significantly decreased with operation time.

As with the $V_{OC}$ (Fig. 8b), there are two decay mechanisms present in the $J_{SC}$ (Fig. 8a). It is known that metal ions diffusing into the organic semiconductors could decrease the $J_{SC}$, as evidenced in studies with Ag nanoparticles in PEDOT:PSS.[162] However, in the present study we see that PFI inhibits the $J_{SC}$ initial fast decay, which, as discussed in the context of *FF*, is not consistent with metal ion diffusion, especially when considering the involved time scales. Metal ion diffusion can then be ruled out to explain the device lifetime alteration in PEDOT:PSS:PFI based devices. Let us consider further that the early degradation mechanism is slower for the devices where PFI is present. Therefore, PFI does not prevent the mechanism responsible for the decay completely or indefinitely.

The most likely explanation is that the hydrophobic property of the fluorinated chains confirmed in Fig. 4 inhibits the diffusion of $H_2O$ to and from the HEL conducting polymer. PEDOT:PSS is relatively more hygroscopic and exposure to $H_2O$ is known to alter the morphology, work-function and conductivity.[53,87] Interestingly, the hydration process does appear to be reversible,[53] and this can explain the PEDOT:PSS work-function significant spread in the literature.[53,65,67] Moreover any changes of HEL hydration mechanism, for instance through PFI blend preparation as herein reported, would obviously alter OPV performances. The encapsulated device, shown in Fig. S14, supports the aforementioned argument.† As with the PFI devices, the degradation of the encapsulated device has gone from being dominated by the $J_{SC}$ to being dominated by the $V_{OC}$ and *FF* when compared with pristine PEDOT:PSS based unencapsulated devices. The percentage decreases in $V_{OC}$ and *FF* between those devices with PFI and the encapsulated devices are similar. The $J_{SC}$ decrease is slightly faster with PFI than with the encapsulated device. This is understandable, as encapsulation with a UV activated optical adhesive and glass cover slip is certain to form a superior $H_2O$ barrier than the PFI alone.

The decreases in *FF* and $V_{OC}$ over time are not significant considering that in the first half-hour the *PCE* of 1:6:0 fell by about 27 %. The majority of which arises from the ~24 % decrease in $J_{SC}$ as shown when comparing Fig. 8c with Fig. 8b. That said, the significant $J_{SC}$ decrease is responsible for the device performance initial degradation, the *FF* and $V_{OC}$ decreases eventually come into play. The resulting *PCE* deterioration for the first half-hour of operation is ~7 % for 1:6:30, i.e. *FF* and $V_{OC}$ account for two thirds of the *PCE* degradation in PFI based devices. This is larger but comparable to the ~ 2 % *PCE* decrease of the encapsulated device. The slowdown in degradation resulting from the PFI is so much so that after 24 hours the *PCE* is still 30 % larger than that of devices made with the pristine HEL.

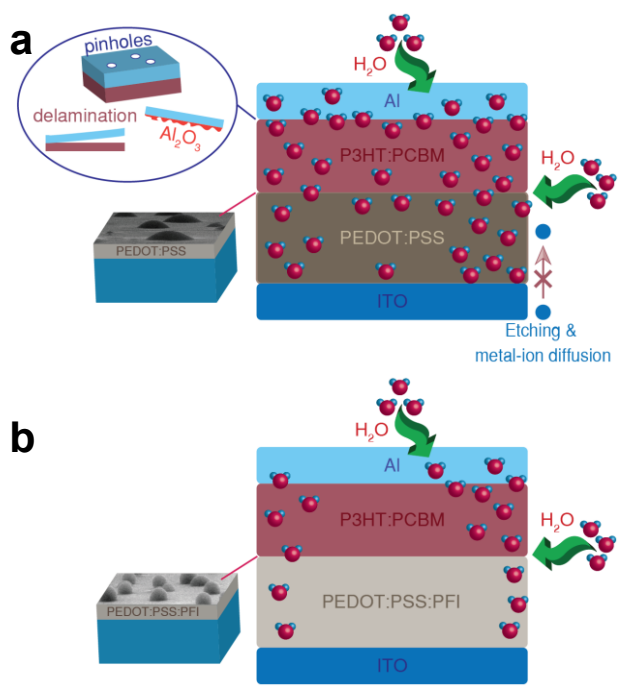

**Fig. 9.** Moisture driven OPV degradation mechanisms, SEM supported illustration of water wetting properties and moisture penetration, and diffusion to the Al electrode on pristine (**a**) and PFI modified (**b**) HEL layer in a P3HT:PCBM device with a standard architecture.

These measurements show that for P3HT:PCBM OPVs, it is the PEDOT:PSS layer that is largely responsible for the burn-in of un-encapsulated devices, i.e. the fast device degradation mechanism is related to the hygroscopic nature of the HEL. We stress that the acid nature of the HEL is not in it-self a concern as by adding PFI molecules in the HEL, the number of sulphonic acid groups is increased, whilst still providing higher *PCE* and slower device degradation. In the HEL herein investigated, the PFI prevents $H_2O$ molecules diffusing to and from the HEL, and altering the properties thereof. Noticeably, altering water molecules mobility could also neutralise the sulphonic group contribution to the device degradation. The mechanisms that are usually seen as responsible for the decrease in $V_{OC}$ and *FF* concern the active layer and more importantly the back electrode. Fig. 9 summarises the degradation pathways present in an un-encapsulated P3HT:PCBM organic photovoltaic cell with a regular non-inverted architecture, and illustrates the main finding of this study. All these pathways can apply to other OPV devices. The degradation at the Al electrode can occur by means of oxidation, delamination, metal ion diffusion and interfacial chemistry among other,[40-44,51] as illustrated on the top left hand-side of Fig. 9.

Indeed, McGehee and co-workers showed that, by replacing the back electrode, one could almost recover the entire loss of $V_{OC}$ and *FF* for P3HT:PCBM photovoltaic cells.[43] In their study, a small $J_{SC}$ decrease was observed, but the OPVs were annealed continuously during the measurements, which would have prevented $H_2O$ from altering the properties of the HEL. Rutherford backscattering (RBS) measurements have also shown that residual water content in the HEL can increase the rate of metal ion diffusion into the PEDOT:PSS.[50] However, this process usually takes place over hours, even if some metal ions can diffuse into the HEL during the spin-coating process. The burn-in of these cells occurs in the first half-hour of operation and is a result of exposure to moisture. The stability of the photoactive materials when exposed to moisture has been confirmed by Poh *et. al* and co-workers.[133] Beyond the photoactive material crystallinity,[165] it is the PEDOT:PSS that is largely responsible for the degradation of such cells.[24] Feron and co-workers demonstrated the ingress of $H_2O$ at the edges of the cell and through pinholes of the aluminium cathode.[26] Furthermore, in their study the degradation was shown to be faster at the edges than through large pin-holes except when the cathode extended beyond the boundary of the active cell. Therefore, not only the PEDOT:PSS degrades itself,[29] even though this degradation appears to be reduced in the case of fluorinated carbon chains carrying the sulphonic acid groups,[100] but the rapid uptake of water also promotes the Al cathode degradation by means of oxidation and/or delamination. This can cause local complete inhibition of charge extraction and it decreases the area available for series conductance.[104]

The results herein presented are then also consistent with the literature showing that the OPV efficiency is sensitive to the aluminium thickness and/or deposition rate.[25] As illustrated in Fig. 4, the use of PFI in PEDOT:PSS prevents water from diffusing to and from the HEL, and hence inhibits the degradation thereof and oxidation and/or delamination of the Al cathode. Interestingly and despite the P3HT hydrophobic nature,[133,166] the photoactive blend does not act as an efficient barrier against moisture diffusion, likely because of both its thickness, the presence of the hygroscopic and interconnected structure of the bulk heterojunction providing a path for moisture to diffuse.[167] Our photovoltaic cells with PFI behave similarly as encapsulated and/or inverted ones where the integrity of the electrodes are preserved and the degradation mechanisms are no longer current dominated, but dominated by the $V_{OC}$ and *FF*.[27] This result is however achieved without having to invert the architecture or encapsulate the devices. Further comparison with the literature would nonetheless be challenging. Indeed, this work looks at fast water induced degradation which we show can largely be prevented. However, a range of mechanisms is involved and can dominate in the burn-in literature. These include slower degradation processes revealed for instance in encapsulated devices where water cannot enter the characterisation cells.

## Conclusions

PEDOT:PSS is known to influence the degradation of organic photovoltaic (OPV) devices, leading to intense efforts to replace it. Our results show that by mediating moisture PEDOT:PSS is largely responsible for the burn-in of P3HT:PCBM OPVs and that using polymeric fluorinated additives is a promising strategy to solve hydroscopic HEL limitations. PEDOT:PSS impacts the degradation of organic solar cells by capturing and releasing moisture, which alters the properties of both the HEL and the Al top cathode. By investigating the PFI concentration effects on optical transmission, topography, work-function, surface wettability by contact angle and environmental scanning electron microscopy, we identify [PFI]≈[PSS] as the PFI concentration threshold above which changes in the HEL properties occur. Below this concentration threshold, OPV devices exhibit poor device performance with a noticeable S-kink in the $J$-$V$ curves, resulting from poor charge extraction. Above this threshold, the device performance recovers and even outperforms the pristine HEL based devices. This results from the fluorinated additive allowing the HEL to reach favourable build-in electric field distributions and reduced interfacial charge recombination, i.e. faster charge extraction. Furthermore, we demonstrate that the fluorinated additive can be used to decrease the amplitude of the device burn-in. This is achieved thanks to its superior wetting properties preventing moisture from diffusing to and from the hygroscopic HEL, which protects the HEL and Al top electrode from deteriorating. By inhibiting the hygroscopic properties of PEDOT:PSS, the fluorinated molecule improves OPV reproducibility, stability and lifetime.

From these findings arises an HEL fluorination strategy applicable to a wide range of OPVs devices using PEDOT:PSS. The degradation mechanisms herein reported should be taken into account to help manage hygroscopic properties and to engineer conducting polymer materials that can help extend the lifetime of organic solar cells.

## Acknowledgements

PA thanks the Canon Foundation in Europe for supporting his visits to the RIKEN through a personal Fellowship, and the Organic Semiconductor Centre of the University of St Andrews for access to facilities. IDWS and CTH acknowledge support from the Engineering and Physical Sciences Research Council of the U.K. (grants EP/L012294/1 and CTH's studentship). IDWS also acknowledges support from a Royal Society Wolfson Research Merit Award.

## Electronic Supplementary Information (ESI)

Experimental section and further details on the interfacial, electronic and photovoltaic cell characterisation.
See https://doi.org/ 10.1039/C8TA04098B